# Role of Nuclear Quadrupole Coupling on Decoherence and Relaxation of Central Spins in Quantum Dots

N. A. Sinitsyn$^a$, Yan Li$^b$, S. A. Crooker$^b$, A. Saxena$^a$, and D. L. Smith$^a$

$^a$ *Theoretical Division, Los Alamos National Laboratory, Los Alamos, NM 87545, USA and*
$^b$ *National High Magnetic Field Laboratory, Los Alamos National Laboratory, Los Alamos, NM 87545, USA*
(Dated: October 30, 2018)

Strain-induced gradients of local electric fields in semiconductor quantum dots can couple to the quadrupole moments of nuclear spins. We develop a theory describing the influence of this quadrupolar coupling (QC) on the spin correlators of electron and hole "central" spins localized in such dots. We show that when the QC strength is comparable to or larger than the hyperfine coupling strength between nuclei and the central spin, the relaxation rate of the central spin is strongly enhanced and can be exponential. We demonstrate a good agreement with recent experiments on spin relaxation in hole-doped (In,Ga)As self-assembled quantum dots.

PACS numbers:

The spin of an electron or a hole in a semiconductor quantum dot is the main component of numerous proposed spintronic and quantum computing devices[1]. Spin decoherence and finite spin lifetimes are currently the major factors that limit our ability to control spin states in dots. A single "central" (*i.e.*, electron or hole) spin in a dot interacts via hyperfine coupling with a large number ($10^4 - 10^6$) of nuclear spins. The net effect of this coupling to the nuclear spin bath can be characterized by an effective Overhauser magnetic field $\mathbf{B_n}$ that acts upon the central spin. Within a quantum dot ensemble, each central spin precesses around a different $\mathbf{B_n}$. If $\mathbf{B_n}$ is time-independent, such precession alone cannot lead to complete relaxation of the central spin polarization. This is evidenced from the observation of spin echoes[2] that can be used to cancel the dephasing of central spins in an ensemble of dots with different constant $\mathbf{B_n}$. However, stochastic dynamics of the Overhauser field $\mathbf{B_n}$ induces irreversible relaxation of the central spin and loss of coherence[3,4]. The physics that leads to changes of $\mathbf{B_n}$ and its corresponding influence on central spin relaxation are the subject of considerable theoretical debate[1,4–8].

It was suggested that, at microsecond time scales, the dynamics of the Overhauser field is dominated by hyperfine-mediated *nuclear co-flips*, which originate from unequal strengths of the hyperfine couplings of the central spin to different nuclear spins inside the same dot[4]. Numerical simulations by Al-Hassanieh et al.[1] showed that such co-flips generally lead only to a logarithmically slow central spin relaxation. In contrast, recent experimental studies with hole-doped (In,Ga)As quantum dots reported a nearly ideal Lorentzian shape of the spin noise power spectrum, indicating exponential relaxation of central hole spins rather than a power-law or logarithmic relaxation[10].

Here we show that quadrupolar couplings (QC) of nuclear spins to the strain induced electric field gradients inside typical semiconductor quantum dots can induce relatively fast dynamics of the Overhauser field $\mathbf{B_n}$, and consequently accelerated relaxation of electron and hole spins in weak external fields. Our model directly applies to InGaAs self-assembled quantum dot systems, which are among the most popular platforms for spin memories and qubits[11,12]; however, the model applies generally to all dots composed of quadrupolar-active nuclei. We model such a nuclear spin bath by introducing static fields acting on nuclear spins due to QC, in addition to the hyperfine couplings to the central spin. We numerically compute the dynamics of our model by applying a time-dependent mean field (TDMF) algorithm[1] that allows us to study the relaxation of a central spin coupled to an unpolarized spin bath containing up to ten thousand nuclear spins.

At low temperatures and at time scales shorter than a millisecond, a Hamiltonian that captures central spin dynamics in quantum dots has the following form:

$$\hat{H} = \sum_{i=1}^{N} \left( \gamma_{||}^i \hat{I}_{iz} \hat{S}_z + \gamma_{\perp}^i (\hat{I}_{ix}\hat{S}_x + \hat{I}_{iy}\hat{S}_y) \right) + g_z B_z \hat{S}_z +$$

$$g_x B_x \hat{S}_x + g_y B_y \hat{S}_y + \sum_{i=1}^{N} \frac{\gamma_Q^i}{2} \left( (\hat{\mathbf{I}}_i \cdot \mathbf{n}_i)^2 - \frac{I(I+1)}{3} \right), \quad (1)$$

where $\hat{S}$ and $\hat{I}_i$ stand for spin operators of, respectively, central and nuclear spins; $B_\alpha$ is an applied magnetic field component along the axis $\alpha$; $g_\alpha$ is the corresponding component of the central spin g-factor. Index $i$ runs though all nuclear spins that interact with the central spin. Parameters $\gamma_{||}^i$ and $\gamma_{\perp}^i$ are the out-of-plane (longitudinal) and in-plane (transverse) coupling strengths, respectively, between the central spin and $i$-th nuclear spin. Henceforth we drop index $i$ for coupling strengths when we discuss their typical magnitudes. For electrons, $\gamma_{||}$ and $\gamma_{\perp}$ have similar magnitudes, but $\gamma_{||}$ and $\gamma_{\perp}$ are quite different for holes. For the latter case, the ratio of transverse to out-of-plane couplings, $\beta = \gamma_{\perp}/\gamma_{||}$ varies in different samples in the range[10,13] $\beta \sim 0.1 - 0.7$. Additional coupling terms in the Hamiltonian such as $\sim \hat{S}_z \hat{I}_{ix}$ are allowed but they were estimated to be negligibly small both in electron and in hole-doped dots[13], and we will disregard them. We also disregard the Zeeman coupling between the external field and nuclear spins because we consider only weak external fields, about the size of the Overhauser field (∼ 25 Gauss for an InGaAs hole doped dot[10]).

The last term in (9) describes QC with characteristic strength $\gamma_Q^i$, and coupling anisotropy vector $\bm{n}_i$ for the $i$-th nuclear spin. QC is allowed for nuclei having spin larger than 1/2. QC has previously proved important in experiments on polarized spin bath relaxation in GaAs[14–18]; however, it has been generally disregarded in the context of central spin relaxation with initially unpolarized nuclear spin baths, both in electron[1,5,6] and in hole-doped[13] dots. We believe that this omission cannot be justified except in certain materials, such as Si, that contain predominantly spin-0 or spin-1/2 nuclei. In the widely studied InGaAs dot system, the most abundant indium isotopes $^{115}$In and $^{113}$In have $I = 9/2$, and Ga and As isotopes have $I = 3/2$. According to many studies[14,15,19], $\gamma_Q \sim 2 - 4$ MHz for indium atoms in GaAs at a typical strain of 3-4% inside a dot, which translates for spin 9/2 into a characteristic level splitting $\gamma_c \equiv \gamma_Q |\bm{I}| \sim 10$ MHz. This value is at least an order of magnitude larger than the effective hyperfine coupling $\gamma_{||} \sim 0.1$-$0.5$ MHz in a typical hole-doped quantum dot with $N \sim 10^5$ nuclei[13,20]. Recent NMR studies of InGaAs dots also showed that the directions of QC anisotropy axes $\bm{n}_i$ are strongly non-uniform inside a dot and do not align with the sample growth anisotropy[19]. To include this fact, we will assume that the local anisotropy vector $\bm{n}_i$ for the $i$-th nuclear spin points in a random direction, which is chosen independently for each nuclear spin. Note that this does not exclude arbitrary spatial correlations of different $\bm{n}_i$ inside the dot.

The Hamiltonian (9) belongs to the class of spin bath models, in which noncollinear static fields act on nuclear spins independently of the coupling to the central spin. In order to compare different models of this class, we introduce a parameter $\gamma_c$ that characterizes the typical energy level splitting of nuclear spins by static fields. Our theory shows that this parameter determines all the essential effects of the static fields irrespective of the details of the interactions. This renders our theory applicable to spin baths with different sizes of nuclear spins $I$. Our results extend beyond the Hamiltonian (9). In fact, the *minimal model* of our class of spin baths can be formulated in terms of the central spin problem with only a nuclear spin-1/2 Hamiltonian:

$$\hat{H} = \bm{B} \cdot \hat{\bm{\sigma}} + \sum_{i=1}^{N}[\gamma_{||}^i \hat{\sigma}_z \hat{\sigma}_z^i + \\ + \gamma_{\perp}^i (\hat{\sigma}_x \hat{\sigma}_x^i + \hat{\sigma}_y \hat{\sigma}_y^i) + \gamma_c^i (\hat{\bm{\sigma}}^i \cdot \bm{n}^i)], \quad (2)$$

where the last term mimics the effect of QC, $\hat{\bm{\sigma}}$ is the Pauli operator of the central (electron or hole) spin, $\hat{\sigma}_\alpha^i$ is the $\alpha$-component of the Pauli operator for the $i$-th nuclear spin, and $\gamma_c^i$ corresponds to the size of the characteristic level splitting for the $i$-th nuclear spin with quantization axis $\bm{n}^i$.

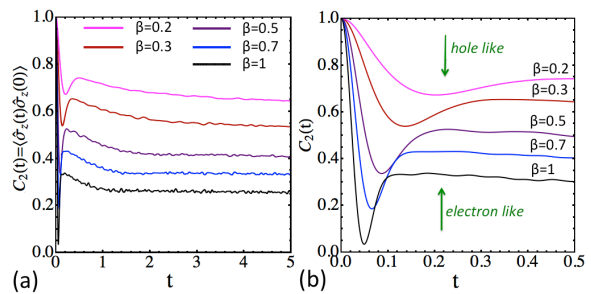

FIG. 1: Spin correlator $C_2(t) = \langle \hat{\sigma}_z(t)\hat{\sigma}_z(0) \rangle$ at $\gamma_c = 0$, $\gamma_{||} \equiv \langle \gamma_{||} \rangle = 1$, $N = 700$ nuclear spins, and $\bm{B} = 0$, shown up to times (a) $t = 5/\gamma_{||}$ and (b) $t = 0.5/\gamma_{||}$ (time $t$ is in units of $1/\gamma_{||}$). Here $\beta = \gamma_\perp/\gamma_{||}$ is the hyperfine coupling anisotropy.

We will compare, in Fig. 2, the dynamics of the model with the Hamiltonian (9) for the spin bath with $I = 1$ and the minimal model (5) at the same characteristic value of $\gamma_c$. Results are almost indistinguishable, so in the rest of the main text, we will show numerical results only for the minimal model to illustrate all the effects.

In the supplementary file[21], we describe the TDMF approach and provide additional numerical tests for evolution with $N$ from 250 to 10000 nuclear spins, the Hamiltonian (9) with $I = 1$, and the classical limit $I \gg 1$, which are all found to be in very good agreement with the theory that we develop here.

Here, we present our results for central spin temporal correlators that were obtained for the model (5) with $N = 700$ spin-1/2 nuclei at equilibrium. Before each simulation, we chose $\gamma_{||}^i = 2\gamma_{||} * r_{1i}$, $\gamma_\perp^i = 2\beta\gamma_{||} * r_{2i}$ and $\gamma_c^i = 2\gamma_c * r_{3i}$, where $r_{1i}$, $r_{2i}$ and $r_{3i}$ are random numbers from a uniform distribution in the interval $(0,1)$. We set the energy scale so that $\gamma_{||} = 1$. Note that we chose widths of parameter distributions to be comparable to the mean values as suggested in [1]. Vectors $\bm{n}^i$ point in random directions and the time step was $dt = 0.0001$. Averaging was performed over 1000 and over 30000 randomly chosen initial state vectors (both for central and nuclear spins) for the calculation of, respectively, noise power and real-time correlators.

In Fig. 13(a) we show our numerical results for the central spin correlator, $C_2(t) = \langle \hat{\sigma}_z(t)\hat{\sigma}_z(0) \rangle$, obtained from the evolution of the Hamiltonian (5) in the absence of quadrupolar interactions ($\gamma_c = 0$). Different curves correspond to different values of the coupling anisotropy $\beta$. All curves start at $C_2(0) = 1$. Figure 13(b) resolves the part of Fig. 13(a) with $t < 0.5$. The appearance and shape of the deep local minimum of $C_2(t)$ in Fig. 13(b) is well understood[4,13] as being due to dephasing caused by ensemble central spin precession around the Overhauser fields $\bm{B_n} = \sum_{i=1}^{N}[\gamma_\perp^i \langle \hat{\sigma}_x^i \rangle \bm{x} + \gamma_\perp^i \langle \hat{\sigma}_y^i \rangle \bm{y} + \gamma_{||}^i \langle \hat{\sigma}_z^i \rangle \bm{z}]$. Figure 13(a) shows that a fraction of the central spin polarization additionally relaxes during a longer time interval that is of order $1/\langle \gamma_{||} \rangle$. This relaxation follows from the co-flip effect[1,5,6]. Figure. 13(a) confirms previ-ous observation[1], which was made for the case $\beta = 1$, that only a fraction of the central spin polarization relaxes via this mechanism on time scales of interest. It also shows that the correlator decay is strongly suppressed by hyperfine coupling anisotropy. This means that standard co-flip effect cannot explain the observed spin relaxation at a fraction of a microsecond in experiments with hole-doped dots[10,11,22], for which $1/\langle\gamma_{||}\rangle$ corresponds to several microseconds[13].

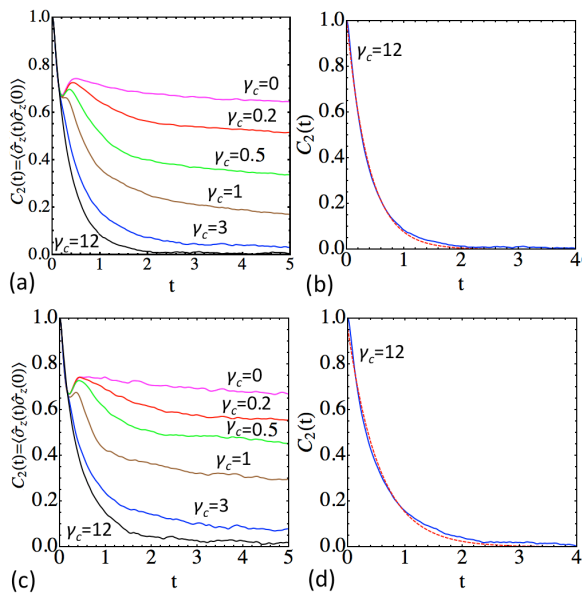

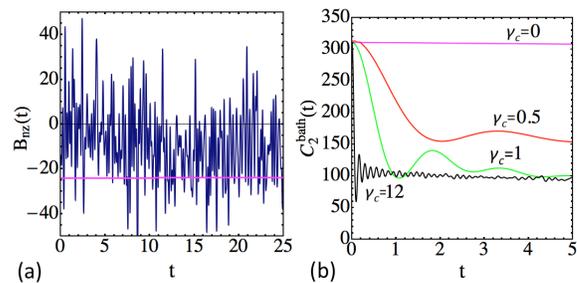

FIG. 2: (a) The real time central spin-spin correlator for different magnitudes of the static field $\gamma_c$ in spin-1/2 bath. (b) Exponential fit (dashed red) of the spin correlator (blue) for spin-1/2 bath at $\gamma_c = 12$ in units of $\gamma_{||}$. (c) Central spin correlator in spin bath with $I = 1$ and quadrupole coupling $\gamma_Q^i = 4\gamma_c^i$. (d) Exponential fit (dashed red) of spin correlator in spin-1 bath at $\gamma_c = 12$. In all cases: $\beta = 0.2$ and $\boldsymbol{B} = 0$.

FIG. 3: For $\beta = 0.2$: (a) Typical Overhauser field dynamics for $\gamma_c = 0$ (pink line) and $\gamma_c = 8$ (blue). (b) Real time Overhauser field correlator $C_2^{\text{bath}}(t) = \langle B_{nz}(t)B_{nz}(0)\rangle$.

Figure 2 shows the central spin correlator $C_2(t)$ for several different mean values of QC [as tuned by the static field $\gamma_c$ to compare spin-1/2 bath in Figs. 2(a-b) and spin-1 bath with $\gamma_Q^i = 4\gamma_c^i$ in Figs. 2(c-d)] at strong anisotropy $\beta = 0.2$. The effect of $\gamma_c \neq 0$ is considerable. Even at $\gamma_c = 0.2 < \langle\gamma_{||}\rangle$, relaxation of the central spin is already much faster than at $\gamma_c = 0$. For $\gamma_c > \gamma_{||}$, we find qualitative changes: the local minimum disappears, relaxation becomes almost complete and furthermore becomes exponential [see Fig. 2(b)]. At $\gamma_c > 8$, the exponential relaxation rate saturates at a value that does not depend on $\gamma_c$ any longer. Figures 2(c-d) show analogous results for the nuclear spin bath with the Hamiltoninan (9) and $I = 1$ with $\gamma_Q^i = 4\gamma_c^i$, which corresponds to the same characteristic splitting of energy levels by QC. It shows that by changing the size of spins and form of the coupling but keeping the same characteristic $\gamma_c$, the form of the central spin correlator does not change.

To better understand the change of behavior with growing $\gamma_c$, it is instructive to look at the dynamics of $B_{nz}$, the Overhauser field component along the $z$-axis. Figure 3 shows examples of $B_{nz}(t)$ starting from a random initial condition for all spins. When $\gamma_c = 0$, $B_{nz}$ is practically frozen. However, for $\gamma_c > \gamma_{||}$ values, $B_{nz}$ quickly fluctuates with the amplitude of the typical Overhauser field strength. Figure 3(b) shows that in the latter case, the bath spin correlator $C_2^{\text{bath}}(t) = \langle B_{nz}(t)B_{nz}(0)\rangle$ decays during $t < 1/\gamma_c$ to a smaller but nonzero value. For $\gamma_c \gg \gamma_{||}$, nuclear spins simply precess around their local static fields from the QC. Fluctuations that are seen in Fig. 3(a) are then merely due to the difference of precession frequencies and precession axis directions for different nuclear spins in one dot. In contrast, when $\gamma_c/\gamma_{||} < 1$, nuclear spin precessions are synchronized by a stronger hyperfine coupling that suppresses fluctuations of $B_{nz}$.

When $\gamma_c > \gamma_{||}$, there can be two distinct regimes of central spin polarization dynamics. The first regime appears when fluctuations of the Overhauser field are so fast that the adiabaticity conditions break down and the central spin polarization cannot follow the direction of the Overhauser field. This most likely can happen when $B_{nz}$ passes through zero values and the Landau-Zener transition probability, $p_{LZ} = 1 - \exp(-\pi(B_{n\perp})^2/v)$ is substantially different from unity. Here $v = (dB_{nz}/dt)_{B_{nz}=0}$ and $B_{n\perp} \sim \gamma_\perp \sqrt{N}$ is the typical value of the Overhauser field transverse to the $z$-axis direction. In this case, each time $B_{nz}$ changes sign, the central spin has substantial probability of not following the Overhauser field so that its dynamics become stochastic with exponential relaxation of the central spin correlator[23]. To estimate $p_{LZ}$ we note that, according to Fig. 3(b), when $\gamma_c > \gamma_{||}$, the time $1/\gamma_c$ sets the scale for the correlator decay time of the Overhauser field. The latter changes during this time by the amount $\delta B_{nz} \sim \gamma_{||}\sqrt{N}$. Hence, the rate of change of the Overhauser field is $v \sim \gamma_{||}\sqrt{N}\gamma_c$, and exponential relaxation can occur when two conditions are satisfied:

$$\gamma_c > \gamma_{||}, \quad \text{and} \quad \eta \equiv \beta^2 \gamma_{||}\sqrt{N}/\gamma_c < 1. \quad (3)$$

For our numerical test with $N = 700$ and $\beta = 0.2$, we find that (19) is satisfied when $\gamma_c/\gamma_{||} \sim 1$. This result is in agreement with Fig. 2. The exponential relaxation time, $\tau_{\text{rel}}$, roughly corresponds to the value of $1/\gamma_c$ at which

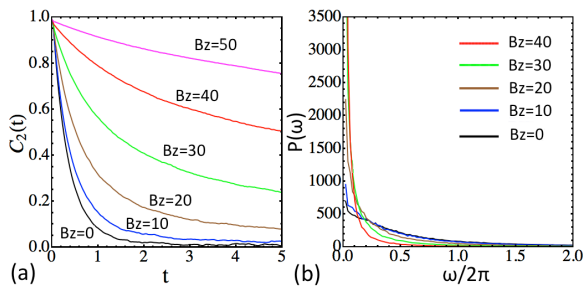

FIG. 4: (a) Real time spin correlator and (b) frequency power spectrum $P(\omega) = \int dt e^{i\omega t} \langle \hat{\sigma}_z(t)\hat{\sigma}_z(0) \rangle$ for different values of external out-of-plane magnetic field $B_z$; $\gamma_c = 8$, and $\beta = 0.2$.

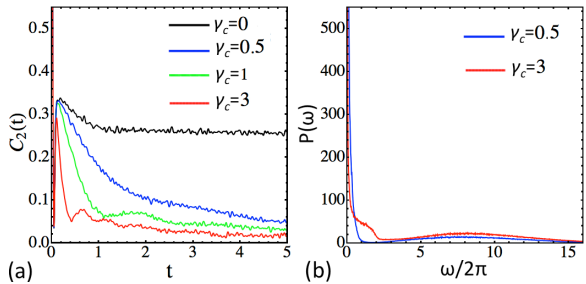

FIG. 5: Central spin correlator at $\beta = 1$ in zero external field: (a) Real time correlator and (b) frequency power spectrum.

$p_{LZ} \sim 1/2$, i.e. $\tau_{\rm rel} \sim 1/[\beta^2 \gamma_{||} \sqrt{N}]$. For the hole-doped dots[10,22], we assume $N = 10^5$, $\beta = 0.2$, $\gamma_c/\gamma_{||} = 25$, which gives $\eta = 0.4 < 1$, i.e. it agrees with the observed Lorentzian shape of the hole spin noise power spectrum in [10]. Considering that $1/\gamma_{||}$ corresponds to several microseconds in hole-doped dots, we find the relaxation time to be a fraction of a microsecond, which also agrees with the experimentally measured value $\tau_{\rm rel} \sim 0.4$ $\mu$s at a zero external field[10]. Our model is also in good agreement with other experimental observations: For example, when an external out-of-plane magnetic field was applied, the central spin relaxation was suppressed[10,11] when this field exceeded $\gamma_{||}\sqrt{N}$. In Fig. 4(a) we confirm this fact numerically. Figure 4(b) also shows our numerical results for the effect of an applied magnetic field on the hole spin noise power spectrum, $P(\omega) = \int dt e^{i\omega t} \langle \hat{\sigma}_z(t)\hat{\sigma}_z(0) \rangle$, which is in good agreement with experimental measurements of this spectrum in external fields[10,11].

The second regime corresponds to the case when fluctuations of the Overhauser field are strong but the central spin follows the direction of the Overhauser field adiabatically. This happens when

$$\gamma_c > \gamma_{||}, \quad \eta > 1. \qquad (4)$$

For *electron-doped* InGaAs dots, condition (4) would likely be satisfied because of a lack of anisotropy ($\beta = 1$). Assuming that such a dot has $N \sim 10^5$ nuclei and $\gamma_c/\gamma_{||} = 3$, we find $\eta \sim 10^2 \gg 1$, i.e. the central spin dynamics is well within the adiabatic regime. For such conditions, the central correlator has to follow the correlation pattern of the Overhauser field, as in Fig. 3(b).

Figure 5(a) shows our results for the real time correlator for $\beta = 1$ (electron-doped dots) and $N = 700$. The case $\gamma_c = 3$ corresponds to conditions (4). The first minimum of $C_2(t)$ in Fig. 5(a) is due to dephasing effects. Note that it is not destroyed by Overhauser field fluctuations, unlike the case with $\beta = 0.2$. At longer times, $C_2(t)$ qualitatively follows the Overhauser field correlation pattern, i.e. it decays to a small but non-zero value during a time $\sim 1/\gamma_c$, followed by a long relaxation tail. Figure 5(b) shows that a specific feature of the regime (4), which distinguishes it from the case with $\gamma_c < \gamma_{||}$, is the appearance of a shoulder in the low frequency peak of the spin noise. An additional feature of the power spectrum at $\eta > 1$ is the presence of a second broad small amplitude Gaussian peak at high frequencies.

*In conclusion,* we identified three regimes with distinct central spin dynamics in the presence of QC at low temperatures and weak external fields: (i) the regime of exponential relaxation of the spin correlator, which is defined by Eq. (19); (ii) the regime with the central spin following Overhauser field adiabatically, which is defined by Eq. (4); and (iii) the regime of weak QC, $0 < \gamma_c < \gamma_{||}$, which is qualitatively similar to $\gamma_c = 0$. We showed that hole-doped InGaAs dots[10] likely correspond to the exponential relaxation regime and that electron-doped dots correspond to the regime (ii). Regime (iii) is potentially applicable to electrostatically defined dots with a nearly perfect atomic lattice.

# Supplementary Material

In this Supplementary we discuss our numerical approaches in more detail and provide representative results for numerical simulations of the central spin problem for larger spin bath sizes and higher spins (with $I = 1$). We discuss our alternative numerical approach, which is based on solving classical Landau-Lifshitz equations with quadratic anisotropies for nuclear spins. We show that although physically justified changes of the model Hamiltonian may lead to observable quantitative differences, our basic conclusions about the role and effects of the quadrupole couplings are robust.

## I. NUMERICAL FRAMEWORKS FOR MODELING CENTRAL SPIN RELAXATION

### A. TDMF framework to simulate static field model of a spin-1/2 spin bath

In this model, we assume that all spins have size 1/2 and then we mimic the effect of QC by introducing random quenched magnetic fields acting on each nuclear spin, so that the full effective Hamiltonian for numerical simulations reads:

$$\hat{H} = \boldsymbol{B}\hat{\boldsymbol{\sigma}}^0 + \sum_{i=1}^{N}(\gamma_{||}^i \hat{\sigma}_z^0 \hat{\sigma}_z^i + \gamma_{\perp}^i (\hat{\sigma}_x^0 \hat{\sigma}_x^i + \hat{\sigma}_y^0 \hat{\sigma}_y^i) + \gamma_c^i (\hat{\boldsymbol{\sigma}}^i \boldsymbol{n}^i)), \quad (5)$$

where the index $i = 0$ stands for the central spin, $\hat{\sigma}_\alpha^i$ is the $\alpha$-component of the Pauli operator for the $i$-th nuclear spin. The vector $\boldsymbol{n}^i$ points in a random direction and $\gamma_c^i$ is the strength of the quenched field acting on the $i$-th nuclear spin. We will assume that $\gamma_c^i$ have the uniform distribution in the interval $(0, 2\gamma_c)$, i.e. with mean $\langle \gamma_c^i \rangle \equiv \gamma_c$. Formally, the substitution of the QC by the random quenched magnetic field changes symmetry properties of this coupling. However, we believe that the Hamiltonian (5) captures the basic qualitative semiclassical picture that leads to the enhanced central spin relaxation. The main advantage of this approach is that there is an efficient well-scalable numerical algorithm, based on the time-dependent mean field (TDMF) dynamics, to simulate the quantum mechanical evolution of the central spin relaxation[1]. Although it is not exact by construction, the TDMF algorithm showed results for the central spin relaxation that are *indistinguishable* from exact quantum mechanical simulations when the number of nuclear spins is relatively large $N > 20$ and when the evolution is averaged over a random nuclear spin distribution at high temperature of the nuclear spin bath[1]. An additional advantage of this model is that the evolution of spins-1/2 in a magnetic field can be relatively quickly computed so that it is possible to simulate the dynamics with a sufficiently large size of the nuclear spin bath.

Following the TDMF theory, we approximate the state vector $|\Psi\rangle$ of the total system as a product $|\Psi\rangle = |u_0\rangle \prod_{i=1}^{N} |u_k\rangle$ of the single-spin vectors $|u_j\rangle$, $(j = 0, \ldots N)$. Then, at each time step we update the state of each spin by considering its evolution with the effective Hamiltonian

$$H_{\text{eff}}^i = \boldsymbol{h}_i(t)\hat{\boldsymbol{\sigma}}^i, \quad i = 0, 1, \ldots, N \tag{6}$$

where the effective field $\boldsymbol{h}_i(t)$ acting on the $i$-th spin is calculated, at each step, according to

$$\boldsymbol{h}_0 = \boldsymbol{B} + \sum_{i=1}^{N} \gamma_{||}^i \sigma_z^i \hat{\boldsymbol{z}} + \gamma_{\perp}^i (\sigma_x^i \hat{\boldsymbol{x}} + \sigma_y^i \hat{\boldsymbol{y}}), \tag{7}$$

$$\boldsymbol{h}_i = \gamma_{||}^i \sigma_z^0 \hat{\boldsymbol{z}} + \gamma_{\perp}^i (\sigma_x^0 \hat{\boldsymbol{x}} + \sigma_y^0 \hat{\boldsymbol{y}}) + \gamma_c^i \boldsymbol{n}^i, \tag{8}$$

where we defined $\boldsymbol{\sigma}^j(t) = \text{Tr}[\hat{\rho}(t)\hat{\boldsymbol{\sigma}}^j]$, and where $\hat{\rho}(t)$ is the density matrix that corresponds to the pure state $\Psi(t)$.

More information about the theoretical justification of this approach can be found in Ref. [1]. Although this reference does not discuss effects of additional quenched fields, one can easily trace that the same steps lead to (6)-(8) when all spins have a value $1/2$.

### B. TDMF framework to simulate spin bath with spins $I > 1/2$ with quadrupole coupling

To simulate dynamics with the Hamiltonian

$$\hat{H} = \sum_{i=1}^{N} \left( \gamma_{||}^i \hat{I}_{iz} \hat{S}_z + \gamma_{\perp}^i (\hat{I}_{ix}\hat{S}_x + \hat{I}_{iy}\hat{S}_y) \right) + g_z B_z \hat{S}_z + \\ + g_x B_x \hat{S}_x + g_y B_y \hat{S}_y + \sum_{i=1}^{N} \frac{\gamma_Q^i}{2} \left( (\hat{\boldsymbol{I}}_i \cdot \boldsymbol{n}_i)^2 \right), \tag{9}$$

with $I > 1/2$, we make a similar mean field approximation that the central spin moves in the effective time-dependent Overhauser field:

$$H_{\text{eff}}^0 = \boldsymbol{h}_0(t)\hat{\boldsymbol{\sigma}}^0, \tag{10}$$

where

$$\boldsymbol{h}_0 = \boldsymbol{B} + \sum_{i=1}^{N} \gamma_{||}^i \hat{I}_z^i \hat{\boldsymbol{z}} + \gamma_{\perp}^i (\hat{I}_x^i \hat{\boldsymbol{x}} + \hat{I}_y^i \hat{\boldsymbol{y}}). \tag{11}$$

There is, however, a complication for nuclear spins because in the spirit of the TDMF-approach, only the coupling to the central spin should be treated in the mean field approximation, while all local fields should be included exactly in numerical simulations. Therefore, the effective Hamiltonian for the nuclear spins is given by

$$\hat{H}_{\text{eff}}^i = \hat{H}_c^i(t) + \hat{H}_Q^i = \boldsymbol{h}_i(t) \cdot \hat{\boldsymbol{I}}^i + \frac{\gamma_Q^i}{2}\left((\hat{\boldsymbol{I}}_i \cdot \boldsymbol{n}_i)^2\right),$$

where $i = 1, 2, \ldots, N$, and where

$$\boldsymbol{h}_i = \gamma_{||}^i \sigma_z^0 \hat{\boldsymbol{z}} + \gamma_{\perp}^i (\sigma_x^0 \hat{\boldsymbol{x}} + \sigma_y^0 \hat{\boldsymbol{y}}). \tag{12}$$

We split the effective Hamiltonian for the $i$-th nuclear spin into the part $\hat{H}_c^i$ that is responsible for interaction with the central spin and the part $\hat{H}_Q^i$ that is responsible for the quadrupole coupling. The reason for this splitting is that, in numerical simulations, we should update the state vector by the evolution operator

$$\hat{U} = e^{i\hat{H}_{\text{eff}}^i dt} \tag{13}$$

at each time step $dt$. For spins $I > 1/2$ this matrix exponent cannot be written explicitly in a compact form. Instead, to perform simulations, e.g. with $I = 1$, we used the Suzuki-Trotter decomposition[2]:

$$\hat{U} \approx e^{i\hat{H}_Q^i dt/2} e^{i\hat{H}_c^i dt} e^{i\hat{H}_Q^i dt/2}, \tag{14}$$

which is the unitary approximation that is correct up to the order $O(dt^2)$. For $I = 1$ and $I = 3/2$, all matrix exponents in (14) can be readily written explicitly.

### C. Spin bath model with classical Landau-Lifshitz dynamics

For spins $I > 3/2$, matrix exponent of $\hat{H}_Q^i$ is not available in analytical form, however, one can apply another approximation $\hat{U} \approx (1+i\hat{H}_{\text{eff}}^i dt/2)(1-i\hat{H}_{\text{eff}}^i dt/2)^{-1}$ which is again unitary and exact up to $O(dt)$. It reduces the calculation of the evolution operator to calculations of the matrix inversion which is readily available. Instead of this approach, however, we explored the case of spin bath dynamics with large nuclear spins by assuming a classical approximation of Landau-Lifshitz dynamics with the Hamiltonian (9).

We considered the *classical* spin Hamiltonian, which correctly takes into account the form of the quadrupole coupling:

$$\hat{H} = \boldsymbol{B}\boldsymbol{S} + \sum_{i=1}^{N}[\gamma_{||}^i S_z s_z^i + \\ + \gamma_{\perp}^i (S_x s_x^i + S_y s_y^i) + \frac{\gamma_Q^i}{2}(\boldsymbol{s}^i \boldsymbol{n}^i)^2] \tag{15}$$

leading to a standard dissipationless Landau-Lifshitz dynamics:

$$\dot{\boldsymbol{s}}_j = \boldsymbol{h}_j \times \boldsymbol{s}_j, \quad \dot{\boldsymbol{S}} = \boldsymbol{h}_0 \times \boldsymbol{S}, \tag{16}$$

$$\boldsymbol{h}_0 = \boldsymbol{B} + \sum_{i=1}^{N} \gamma_{||}^i s_z^i \hat{\boldsymbol{z}} + \gamma_{\perp}^i (s_x^i \hat{\boldsymbol{x}} + s_y^i \hat{\boldsymbol{y}}), \tag{17}$$

$$\boldsymbol{h}_i = g_{||}^i S_z \hat{\boldsymbol{z}} + \gamma_{\perp}^i (S_x \hat{\boldsymbol{x}} + S_y \hat{\boldsymbol{y}}) + \gamma_Q^i (\boldsymbol{s}^i \boldsymbol{n}^i)\boldsymbol{n}^i. \tag{18}$$

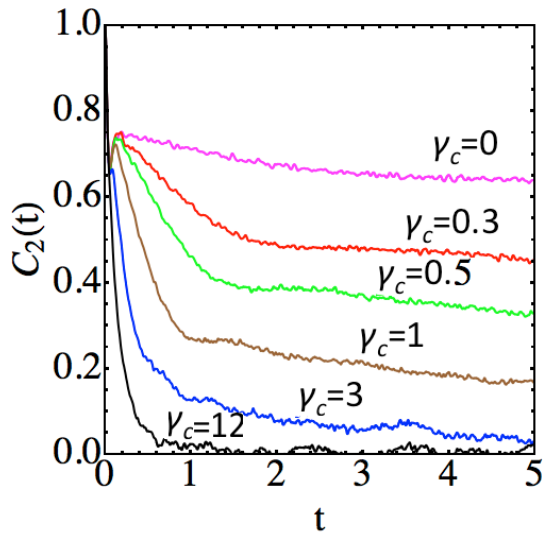

FIG. 6: The real time spin-spin correlator for nuclear spins baths with the Hamiltonian (5) and $N = 10000$ nuclear spins; $\beta = 0.2$, and $\boldsymbol{B} = 0$. Different curves correspond to different values of $\gamma_c$. Time is in units of $1/\gamma_{||}$. Averaging is over 5000 runs with different random initial conditions.

The classical treatment of nuclear spins should be additionally justified in our case by large sizes of nuclear spins (9/2 for indium isotopes). For simplicity, we will normalize to one all spin sizes in this classical model. The advantage of the classical approach is its speed. It also correctly preserves all symmetries of the couplings. Its disadvantage is its inability to take rigorously into account the discreteness of spin states, which turned out to be not an important factor to obtain qualitative behavior of the central spin problem.

## II. SCALING OF CENTRAL SPIN CORRELATORS WITH THE NUMBER $N$ OF NUCLEAR SPINS

As an additional test of our theory, we performed a number of simulations of the central spin dynamics with different sizes of the nuclear spin bath in the range from $N = 250$ to $N = 10000$ nuclear spins. To obtain results with bigger $N$ we enhanced precision by decreasing the size of the time step in simulations to $dt = 0.00005$ and reducing the statistics of averaging over different random initial conditions from 20000 runs for $N < 1000$ to 5000 runs at $N = 10000$. In the main text, we made two analytical predictions which are particularly suitable to test by changing $N$. First, we predict that the transition between the exponential regime and the regime at which the central spin polarization follows the Overhauser field adiabatically happens in the region of parameter values, at which the following condition is satisfied:

$$\eta \equiv \beta^2 \gamma_{||} \sqrt{N}/\gamma_c \sim 1, \qquad (19)$$

with the exponential relaxation regime corresponding to $\eta < 1$.

Figure 6(a) shows the effect of increasing $\gamma_c$ for spin bath with $N = 10000$ nuclear spins. At $\gamma_{||} = 1$ and $\beta = 0.2$, condition (19) would be satisfied for $\gamma_c \approx 4$. Indeed, Fig. 6(a) shows that near the value of $\gamma_c = 3$ the local dephasing minimum at short time-scales is still present but almost disappeared. At $\gamma_c = 12$ the relaxation is exponential. One can compare this figure with analogous Fig. 2 in the main text for $N = 700$. In the latter case, similar disappearance of the dephasing minimum of $C_2(t)$ was observed at a smaller value of $\gamma_c \sim 1$, which is in agreement with the $\sqrt{N}$ factor in (19).

Another apparent difference of cases with $N = 700$ and $N = 10000$ is that at the larger $N$ the local dephasing minimum of $C_2(t)$ shifts toward the shorter time scales. This is expected because this minimum corresponds to the typical rotation time of the central spin around the Overhauser field. This field increases with $N$ as $\sqrt{N}$, which explains the shift of the local minimum. We note also that increasing $N$ does not change our conclusion about the absence of substantial relaxation of the central spin at $\gamma_c = 0$ during times longer than $1/\gamma_{||}$.

Our second analytical prediction states that in the exponential regime of relaxation, the relaxation rate is given by

$$\tau_{\rm rel} = \frac{A}{\beta^2 \gamma_{||} \sqrt{N}}, \qquad (20)$$

where $\beta$ is the anisotropy factor, $\gamma_{||}$ is the typical strength of hyperfine coupling and $N$ is the number of spins. Coefficient $A$ is a numerical factor, which is not universal and may depend on the specific characteristics of the nuclear spins, for example the sizes of the nuclear spins and the distribution of parameters. Nevertheless, considering the generality of the arguments leading to Eq. (20), this coefficient is expected to be of the order unity.

In order to test prediction (20), we considered a nuclear spins-1/2 bath with anisotropy $\beta = 0.2$ and $\gamma_c/\gamma_{||} = 12$. According to (19) the condition for exponential relaxation, $\eta < 1$, is satisfied at $\gamma_c = 12$ and $\beta = 0.2$ for nuclear spin baths with up to $N \sim 90000$ nuclear spins. Figure 7 shows our numerical results for the relaxation times of correlators that were obtained for different $N$ by the best exponential fits of the correlator relaxation curves (blue boxes). Red curve shows the best fit of these numerically obtained points by Eq. (20) with a single free fitting parameter $A$. The best fit was obtained for $A \approx 0.45$, which is of the order unity, as expected. Figure 7 demonstrates quite a reasonable agreement with prediction (20).

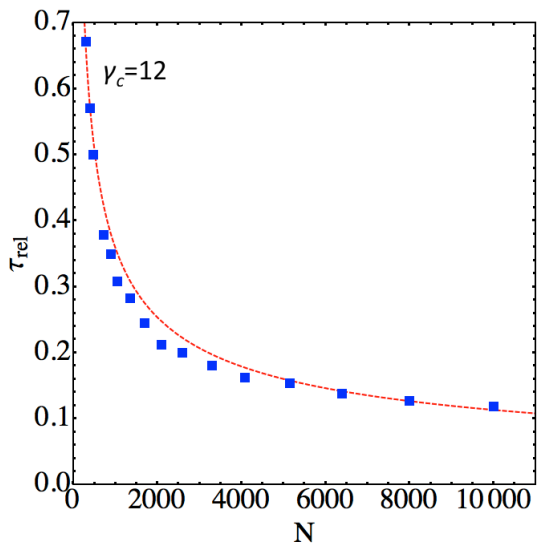

FIG. 7: Relaxation time as a function of the number $N$ of nuclear spins-1/2 in the bath. $\gamma_{||} = 1$, $\beta = 0.2$, $\gamma_c = 12$, which corresponds to the exponential regime. Time is in units of $1/\gamma_{||}$. Blue boxes correspond to results of numerical simulations and dashed red curve is given by Eq. (20) with the numerical factor $A \approx 0.45$.

## III. CENTRAL SPIN CORRELATORS WITHIN TDMF APPROACH WITH SPIN-1 NUCLEAR SPINS

### A. Relaxation for different $\gamma_c = \gamma_Q/4$

To show that our results are not specific for nuclear spins-1/2, we performed TDMF simulations for spins-1, treating the quadrupole coupling exactly with 2nd order Suzuki-Trotter decomposition[2], as explained in Section 1.B.

In order to make quantitative comparison with results for spin-1/2, we note that the value $2\gamma_c$ was the splitting of energy levels in the spin-1/2 Hamiltonian, which is induced by static fields. The corresponding term in the Hamiltonian with quadrupole coupling, $(\gamma_Q/2)(\boldsymbol{n}\cdot\hat{\boldsymbol{I}})$ introduces the level splitting of the size $\gamma_Q/2$ between nearest energy levels. Hence, for $|\boldsymbol{I}| = 1$ we introduce parameter $\gamma_c = \gamma_Q/4$, which characterizes the level splitting by quadrupole field. Since we expect that $\gamma_c$ is the main parameter that determines the effect of quadrupole coupling, the behavior of spin correlators at same values of $\gamma_c$ should be similar for spin-1/2 and spin-1 nuclear spin bathes.

Figure 8 shows the results of our simulations for spin-1 nuclear spin bath. It reproduces Figs. 2(c,d) of the main text. Distributions of parameters $\gamma_{||}^i$ and $\gamma_{\perp}^i$ were chosen the same as in the case of spin-1/2 bath in the main text. Quadrupole coupling for i-th nuclear spin was defined $\gamma_Q^i = 4\gamma_c^i$, where $\gamma_c^i$ was chosen according to the same distribution as in the main text for the spin-1/2 bath.

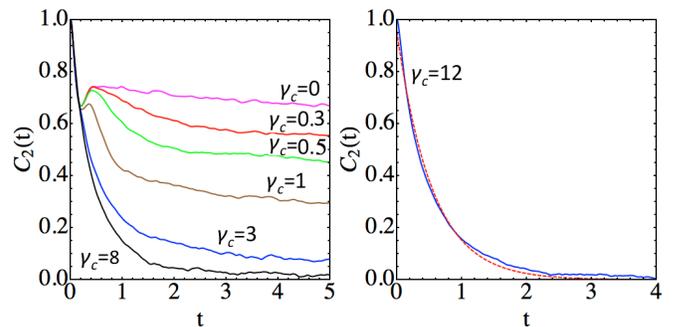

FIG. 8: (a) The real time spin-spin correlator (blue curve) for nuclear spin $I = 1$ baths with different magnitudes of the quadrupolar coupling. We introduce a characteristic level splitting parameter $\gamma_c \equiv \gamma_Q/4$ in oder to compare predictions of spin-1/2 and spin-1 baths; $\beta = 0.2$, $N = 700$, and $\boldsymbol{B} = 0$. (b) Exponential fit (dashed red) of the spin correlator at $\gamma_c = 12$ in units of $\gamma_{||}$ (red). Time is in units of $1/\gamma_{||}$.

Comparing Figs. 2(c,d) and Figs. 2(a,b) of the main text, which describe, respectively spin-1 and spin-1/2 baths, we find that the differences between them are not essential despite difference of the nuclear spin sizes and the form of the considered nuclear spin coupling to strain fields. As in the analogous example of spin-1/2, the transition to the exponential regime happens at a characteristic parameter value $\gamma_c \sim 1$, with relaxation exponent saturating near $\gamma_c \sim 8$. More detailed examination reveals small differences, for example, for $\gamma_c = 12$, the characteristic relaxation time obtained by the best exponential fit for spin-1 bath is equal to $\tau_{\rm rel}^{I=1} \approx 0.50$, while for the spin-1/2 bath in Fig. 2 of the main text it was $\tau_{\rm rel}^{I=1/2} \approx 0.39$. Such quantitative differences up to a coefficient of order unity are expected and do not affect any of the basic predictions of our theory.

### B. Relaxation at fixed $\gamma_Q$ but different $N$

Figure 9 tests the $1/\sqrt{N}$ prediction for the relaxation time in the exponential regime for spin baths with nuclear spins $I = 1$ and $N$ in the range from $N = 200$ to $N = 4000$. Again, as in the case of spins-1/2, the agreement with theoretical prediction is quite reasonable. Quantitative differences between cases with defferent nuclear spin sizes mainly reduce to a small difference of the corresponding coefficients $A$: $A \approx 0.56$ for spins-1 bath and $A \approx 0.45$ for spins-1/2 bath.

### C. Effect of coupling randomness

Here we use the TDMF approach for $I = 1$ to investigate the role of coupling randomness. In all our previous examples, we assumed that parameters $\gamma_{||}^i$, $\gamma_{\perp}^i$ and $\gamma_c^i$ have a broad distribution with the variance close to the mean value. However, most of our discussion for the transition to exponential regime and the relaxation time value did not involve this randomness assumption.

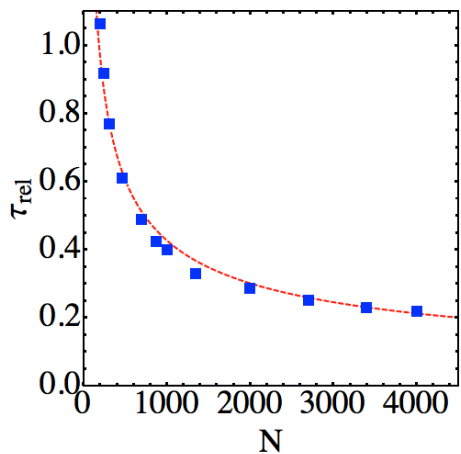

FIG. 9: Relaxation time as function of the number $N$ of nuclear spins with $I = 1$ in the bath. $\gamma_{||} = 1$, $\beta = 0.2$, $\gamma_c = \gamma_Q/4 = 12$, which corresponds to the exponential regime. Time is in units of $1/\gamma_{||}$. Blue boxes correspond to results of numerical simulations and dashed red curve is given by Eq. (20) with the numerical factor $A \approx 0.56$. Averaging is performed over 5000 runs with different initial random conditions for all spins.

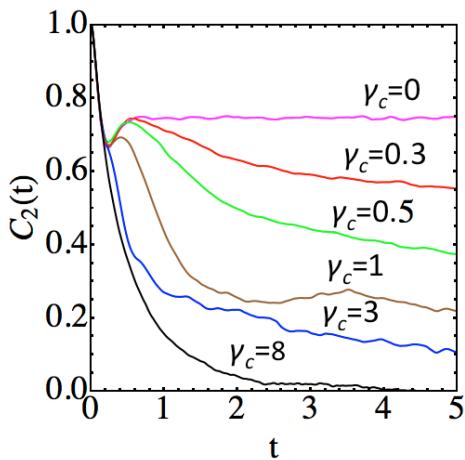

FIG. 10: Spin-spin correlators in the absence of randomness for parameters $\gamma_{||}^i$, $\gamma_\perp^i$ and $\gamma_c^i$. The nuclear spin bath has $I = 1$ with random choice of anisotropy vectors $\boldsymbol{n}^i$. Level splitting parameter is $\gamma_c \equiv \gamma_Q/4$; $\beta = 0.2$, $N = 700$, and $\boldsymbol{B} = 0$. Averaging is over 10000 runs with random initial conditions.

We performed simulations, analogous to the ones in Fig. 8, but assuming that $\gamma_{||}^i = \gamma_{||}$, $\gamma_\perp^i = \gamma_\perp$ and $\gamma_c^i = \gamma_c$ for any $i = 1, \ldots N$. We kept the assumption of random $\boldsymbol{n}^i$ only. While such a spin bath may be considered unphysical, it is instructive to look at its effects in order to illustrate the effect of parameter randomness,

Figure 10 shows results of such simulations. First, we note that the transition to the exponential regime above

$\gamma_c \sim 1$ remains true, and at $\gamma_c = 8$ we find an almost exponential relaxation, as usual. If we compare relaxation rates, we find that they are not influenced by parameter randomness. For example, for the case in Fig. 8 the best exponential fit of $\gamma_c = 8$ curve corresponds to $\tau_{\rm rel} \sim 0.50$ and for the case in Fig. 10 we found $\tau_{\rm rel} \sim 0.51$. This example illustrates the universality of behavior in the exponential regime if it is achieved. While for small and intermediate $\gamma_c$ values, strong changes of parameter distribution, in principle lead to noticeable changes of relaxation pattern, after reaching the exponential regime the differences between cases with different distributions of parameters become practically unnoticeable.

At the regime with $\gamma_c/\gamma_{||} < 0.5$, the above behavior also changes insignificantly. The difference becomes clear if we look at the $\gamma_c = 0$ case. Then the model with identical parameters predicts no relaxation at all after going through the dephasing minimum.

Considerable changes of the behavior are observed at $\gamma_c/\gamma_{||} \sim 1$, for which a very complex relaxation pattern is shown in Fig. 10. This behavior is not surprising either. Indeed, this case corresponds to a situation at which fluctuations of the Overhauser field are substantial but the central spin follows those fluctuations almost adiabatically. For constant parameters in this regime, the correlator of the Overhauser field can show unusually large oscillations, which at normal conditions would be smeared by parameter randomness. The central spin follows that Overhauser field and, in this regime, its correlator reflects oscillating behavior of the Overhauser field. This explains the unusual relaxation at $\gamma_c/\gamma_{||} \sim 1$ in Fig. 10.

## IV. CENTRAL SPIN CORRELATORS IN SEMICLASSICAL LANDAU-LIFSHITZ DYNAMICS

The complexity of simulating spin baths with higher sizes of nuclear spins increases. To get an insight in spin bath effects with large values of spins, we simulated the nuclear spin bath dynamics classically within the Hamiltonian (15). As it was expected from the semiclassical character of our theory, we obtained essentially the same relaxation behavior of the central spin as in the case of the quantum mechanical TDMF framework with quenched magnetic fields. The quantitative differences should be attributed to normalization [$C_2(0) = 1/3$ for classical unit spin rather than $C_2(0) = 1$ for quantum spin-1/2] and to different functional forms of the static field terms assumed in the two frameworks. Since all results that we obtained are qualitatively the same as in the case of the TDMF framework with quenched magnetic fields, we will only show a few examples.

We performed simulations of the model (15) for a central spin coupled to $N = 700$, and in a few instances to $N = 4000$ nuclear spins. We chose quadrupole couplings to be uniformly distributed random numbers $\gamma_Q^i = \gamma_Q * RND_i$, where $[RND_i \in (0, 2)]$, and the direc-tion of the anisotropy axis $\mathbf{n}_i$ was chosen randomly for each nuclear spin.

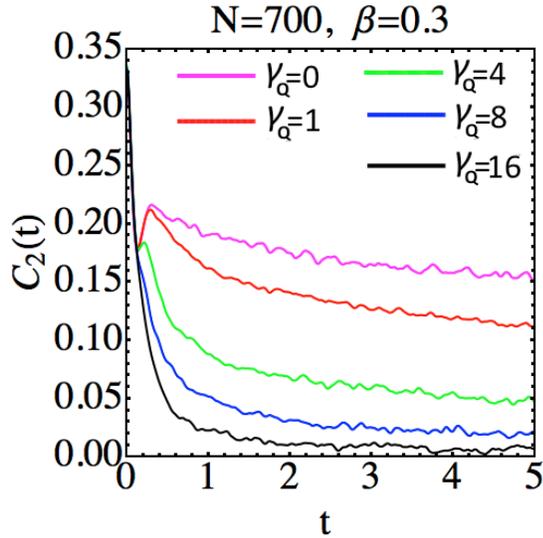

FIG. 11: Real time spin-spin correlator produced by classical Landau-Lifshitz dynamics with Hamiltonian (15). Here $N = 700$, $\langle \gamma_\| \rangle = 1$, $\beta = 0.3$. Results are averaged over 20000 runs with random initial conditions for central and nuclear spins. The correlator of classical spin with $|\mathbf{S}| = 1$ starts at $C_2(0) = 1/3$.

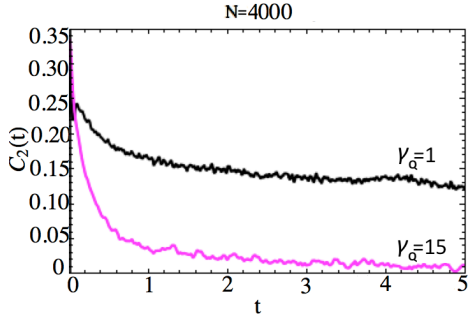

FIG. 12: Real time spin correlator in classical Landau-Lifshitz dynamics with $N = 4000$ nuclear spins, $\langle \gamma_\| \rangle = 1$, $\beta = 0.2$. Results are averaged over 5000 runs with random initial conditions for central and nuclear spins.

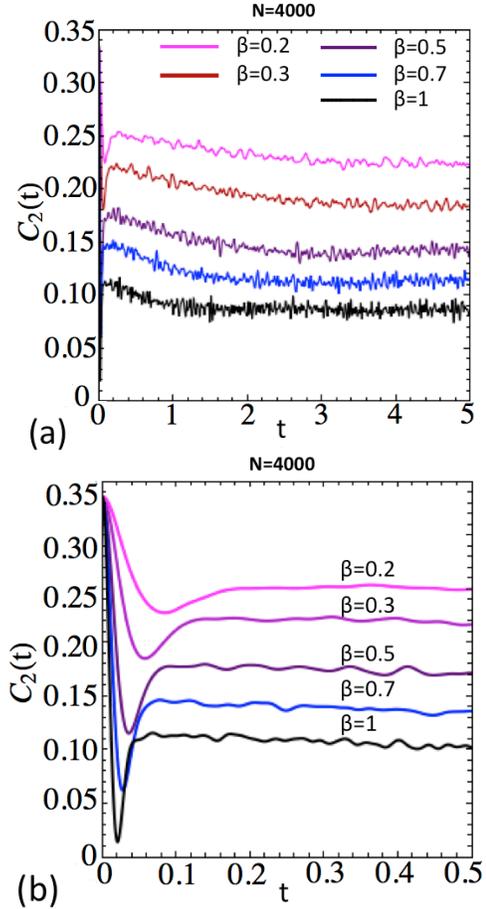

FIG. 13: Real time spin-spin correlator $C_2(t) = \langle S_z(t) S_z(0) \rangle$ in classical Landau-Lifshitz dynamics with the Hamiltonian (15), at $\gamma_Q = 0$, up to times (a) $t = 5/\langle \gamma_\| \rangle$ and (b) $t = 0.5/\langle \gamma_\| \rangle$. Different curves correspond to different sizes of the coupling anisotropy, $\beta = \langle \gamma_\perp \rangle / \langle \gamma_\| \rangle$. Here $\mathbf{B} = 0$, $N = 4000$, $\langle \gamma_\| \rangle = 1$. Averaging is (a) over 5000 and (b) over 25000 simulations with different random initial states of central and nuclear spins, as well as hyperfine couplings.

According to Fig. 11, as in the case of quenched fields, there is a critical value of $\gamma_Q$ above which the central spin relaxation becomes almost exponential and the local minimum due to the coherent central spin precession disappears. To show that this behavior is not specific for $N = 700$ nuclear spins, Fig. 12 shows $C_2(t)$ for a larger number of nuclear spins ($N = 4000$).

Finally, Fig. 13 shows the numerically calculated spin correlator for the model with the Hamiltonian (15) at $\gamma_Q = 0$ with $N = 4000$. Different curves correspond to different values of the coupling anisotropy $\beta$. Figure 13(b) resolves the part of Fig. 13(a) with $t < 0.5$.

## V. EFFECT OF NUCLEAR STATIC FIELD ANISOTROPY

In the main text, we assumed that quenched fields can equally probably point in any direction. In reality, we expect that anisotropy of the lattice is reflected in anisotropy of quadrupole coupling fields. To explore this case within TDMF approach, we considered a cubic anisotropy by choosing random quenched fields $\boldsymbol{\gamma}_c^i$ in spin-1/2 bath having independent random components $\gamma_{cx}^i$, $\gamma_{cy}^i$, and $\gamma_{cz}^i$, each uniformly distributed in the interval $(-\gamma_c, \gamma_c)$. Such random field vectors are uniformly distributed inside a cube rather than a sphere. Figure 14, which is the counterpart of Fig. 3 in the main text, shows that the correlation function of the Overhauser field [Fig. 14(b)] is quite different from the one in Fig. 3 of the main text but the sample of a fluctuating trajectory, shown in Fig. 14(a) is similar to the isotropic case. This means that our basic discussion about the breakdown of adiabaticity should equally apply to the case of cubic anisotropy, leading to the same quantitative predictions. Indeed, Fig. 15, which is the analog of Fig. 2(a,b) in the main text, shows that a cubic anisotropy of quenched fields does not affect our conclusions about relaxation rates at different strengths of $\gamma_c$. Effects of additional in-plane or out-of-plane anisotropy of $\gamma_c$ can be more pronounced because in-plane anisotropy favors stronger fluctuations of $B_{nz}$, while out-of-plane anisotropy suppresses fluctuations of $B_{nz}$.

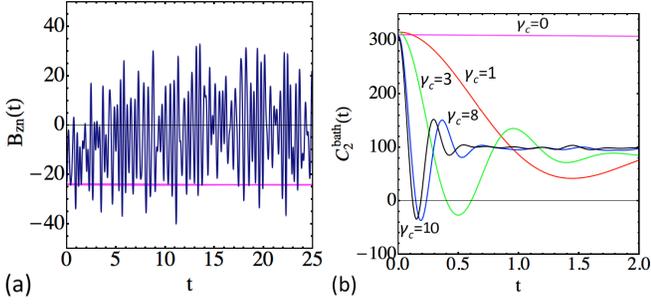

FIG. 14: For $\beta = 0.2$ and cubic anisotropy of $\boldsymbol{\gamma}_c$: (a) Typical Overhauser field dynamics for $\gamma_c = 0$ (pink) and $\gamma_c = 8$ (blue); (b) real time Overhauser field correlator. $N = 700$.

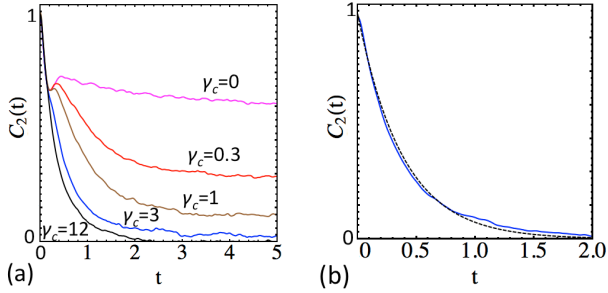

FIG. 15: Spin correlators at cubic anisotropy of $\boldsymbol{\gamma}_c$: (a) Real time spin correlator for different magnitudes of the quenched field $\gamma_c$, $\beta = 0.2$, $\boldsymbol{B} = 0$; (b) exponential fit (dashed black) of spin correlator for $\gamma_c = 8$ (blue). $N = 700$.